\documentclass[11pt, reqno]{amsart}
\usepackage[english]{babel}
\usepackage[table]{xcolor}
\usepackage[utf8]{inputenc}
\usepackage{amsmath, amsfonts, amsthm, amscd, amssymb, fullpage, fancyhdr, upgreek, amssymb, hyperref, enumerate, comment, siunitx, enumitem, graphicx, mathrsfs, mathtools, xcolor,
microtype}
\usepackage{listings}
\usepackage[scr]{rsfso}
\usepackage{thmtools}
\usepackage{thm-restate}

\newcolumntype{P}[1]{>{\centering\arraybackslash}p{#1}}

\usepackage[parfill]{parskip}

\newtheorem{theorem}{Theorem}[section]

\newtheorem{corollary}[theorem]{Corollary}
\newtheorem{lemma}[theorem]{Lemma}
\newtheorem{conjecture}[theorem]{Conjecture}

\theoremstyle{definition}
\newtheorem{definition}[theorem]{Definition}

\newtheorem*{remark}{Remark}

\newcommand{\bburl}[1]{\textcolor{blue}{\url{#1}}}

\usepackage{tikz}
\usepackage{tkz-tab}
\usepackage{tkz-graph}
\usetikzlibrary{shapes.geometric,positioning}
\usetikzlibrary{bayesnet}

\numberwithin{equation}{section}

\newcommand{\hr}[1]{\href{#1}{\url{#1}}}

\newcommand{\Prob}{\mathbb{P}}
\newcommand{\Exp}{\mathbb{E}}

\newcommand{\perc}{\text{Perc}}
\newcommand{\pa}{\text{pa}}

\begin{document}

\title{Strong Data-Processing Inequalities and their Applications to Reliable Computation}

\author[Yang]{Andrew K. Yang}
\email{\textcolor{blue}{\href{mailto:aky30@cantab.ac.uk}{aky30@cantab.ac.uk}},
\textcolor{blue}{\href{mailto:andrewkelvinyang@gmail.com}{andrewkelvinyang@gmail.com}}}
\address{Emmanuel College, University of Cambridge, Cambridge, CB2 3AP, UK}

\thanks{The choice of topics covered in this essay take inspiration from Chapter 33 of Polyanskiy and Wu's textbook \emph{Information Theory: From Coding to Learning} \cite{PWTB}.}

\begin{abstract}
    In 1952, von Neumann gave a series of groundbreaking lectures that proved it was possible for circuits consisting of 3-input majority gates that have a sufficiently small independent probability $\delta > 0$ of malfunctioning to reliably compute Boolean functions. In 1999, Evans and Schulman used a strong data-processing inequality (SDPI) to establish the tightest known necessary condition $\delta < \frac{1}{2} - \frac{1}{2\sqrt{k}}$ for reliable computation when the circuit consists of components that have at most $k$ inputs. In 2017, Polyanskiy and Wu distilled Evans and Schulman's SDPI argument to establish a general result on the contraction of mutual information in Bayesian networks.
    
    In this essay, we will first introduce the problem of reliable computation from unreliable components and establish the existence of noise thresholds. We will then provide an exposition of von Neumann's result with 3-input majority gates and extend it to minority gates. We will then provide an introduction to SDPIs, which have many applications, including in statistical mechanics, portfolio theory, and lower bounds on statistical estimation under privacy constraints. We will then use the introduced material to provide an exposition of Polyanskiy and Wu's 2017 result on Bayesian networks, from which the 1999 result of Evans-Schulman follows.
\end{abstract}

\subjclass[2020]{TBD}


\maketitle

\setcounter{tocdepth}{1}
\tableofcontents

\newpage
\section{Introduction}

In 1938, Shannon proved that any function from $\{0,1\}^n \to \{0,1\}$ (a Boolean function) can be computed by a Boolean circuit constructed with AND, OR and NOT gates \cite{S38}. The individual logic gates used in Shannon's circuits are idealistic - they never fail. von Neumann became interested in whether computation by circuits was possible if the individual components were imperfect and prone to a failing at random with probability $\delta$ (noise). He believed this ought to be the case, as he interpreted the brain as consisting of unreliable components, but also able to reliably perform difficult tasks.

In 1952, von Neumann proved that any Boolean function that can be computed by a circuit of 3-input majority (3MAJ) gates can be reliably computed by a noisy circuit of 3MAJ gates provided the noise is sufficiently small \cite{V52}. We provide a proof of von Neumann's result, and also the proof of a similar result with 3-input minority (3MIN) gates, in Section \ref{sec:von}. We extend the work in this way because this allows us to reliably compute all Boolean functions, rather than a subset of them.

In 1988, Pippenger introduced information theoretic techniques, including the Strong Data-Processing Inequality (SDPI), to find an upper bound on $\delta$ for which reliable computation by Boolean \emph{formulas} (a special type of Boolean circuit) is possible \cite{P88}. A year later, Feder used ideas from percolation theory to generalise Pippenger's result to hold for noisy circuits \cite{F89}. 

Finding the threshold where reliable computation by circuits is possible if and only if $\delta$ is below the threshold remains an open question. In Section \ref{sec:comp}, we establish that such a threshold exists. The best known upper bound on the threshold, which we denote $\delta^*(k)$, is

\begin{equation*}
    \delta^*_{\text{ES}} (k) \ = \ \frac{1}{2} \ - \ \frac{1}{2\sqrt{k}}, 
\end{equation*}

proved by Evans and Schulman in 1999 \cite{ES99} using improvements of ideas from Pippenger's 1998 and Feder's 1999 papers. Here, $k$ is the fan-in of the circuit - the largest number of inputs any gate in $k$ can have. We will prove Evans-Schulman's result in Section \ref{sec:pol}. 

Considerably more is known when we restrict from Boolean circuits to Boolean formulas. In 1991, Hajek and Weller applied a new approach to find the threshold noise for reliable computation by Boolean \emph{formulas} with fan-in 3 to be $\delta^*_f (3) = \frac{1}{6}$. Their work is markedly different to the contents of this essay.

This landmark result was later extended by Evans and Schulman, who found the threshold for all odd $k$ fan-in formulas \cite{ES03}, finding

\begin{equation*}
    \delta^*_f (k) \ = \ \frac{1}{2} \ - \ \frac{2^{k-2}}{k \binom{k-1}{(k-1)/2}}.
\end{equation*}

We use these results about noisy formulas to establish Lemma \ref{lem:converg}, showing that $\delta^*_{\text{ES}} (k)$ is an asymptotically tight upper bound for $\delta^*$ when $k$ is large.

\section{Reliable Computation from Unreliable Components}
\label{sec:comp}

In this section, we will provide the necessary background to the field of reliable computation before we provide a proof of von Neumann's Theorem \ref{thm:von} in the next section. We conclude the section by proving of Lemma \ref{lem:converg}, which allows us to appreciate the tightness of Evans-Schulman's upper bound.

\begin{figure}
    \centering
    \begin{minipage}[t]{0.45\textwidth}
        \centering
        \begin{tikzpicture}[scale=1.1]
            \node[draw] (1) at (0,4) {$x_1$};
            \node[draw] (2) at (1,4) {$x_2$};
            \node[draw] (3) at (2,4) {$x_3$};
            \node[draw] (4) at (3,4) {$x_4$};
            \node[draw] (5) at (4,4) {$x_5$};
            \node[draw] (a) at (0.5,3) {$G_1$};
            \node[draw] (b) at (2.5,3) {$G_2$};
            \node[draw] (c) at (1,2) {$G_3$};
            \node[draw] (d) at (3.5,2) {$G_4$};
            \node[draw] (e) at (2.25,1) {$G_5$};
            \node[] (f) at (2.25,0) {$C(x_1,\dots,x_n)$};
            \draw[->] (1) -- (a);
            \draw[->] (2) -- (a);
            \draw[->] (3) -- (b);
            \draw[->] (4) -- (b);
            \draw[->] (a) -- (c);
            \draw[->] (b) -- (c);
            \draw[->] (b) -- (d);
            \draw[->] (5) -- (d);
            \draw[->] (c) -- (e);
            \draw[->] (d) -- (e);
            \draw[->] (e) -- (f);
        \end{tikzpicture}
    \end{minipage}\hfill
    \begin{minipage}[t]{0.45\textwidth}
        \centering
        \begin{tikzpicture}[scale=1.5]
            \node[draw, circle] (1) at (0,4) {};
            \node[draw, circle] (2) at (1,4) {};
            \node[draw, circle] (3) at (2,4) {};
            \node[draw, circle] (4) at (3,4) {};
            \node[draw, circle] (5) at (4,4) {};
            \node[draw, circle] (a) at (0.5,3) {};
            \node[draw, circle] (b) at (2.5,3) {};
            \node[draw, circle] (c) at (1,2) {};
            \node[draw, circle] (d) at (3.5,2) {};
            \node[draw, circle] (e) at (2.25,1) {};
            \draw[->] (1) -- (a);
            \draw[->] (2) -- (a);
            \draw[->] (3) -- (b);
            \draw[->] (4) -- (b);
            \draw[->] (a) -- (c);
            \draw[->] (b) -- (c);
            \draw[->] (b) -- (d);
            \draw[->] (5) -- (d);
            \draw[->] (c) -- (e);
            \draw[->] (d) -- (e);
        \end{tikzpicture}
    \end{minipage}
    \caption{Example circuit and corresponding DAG}
    \label{fig:booleanCircuit}
\end{figure}
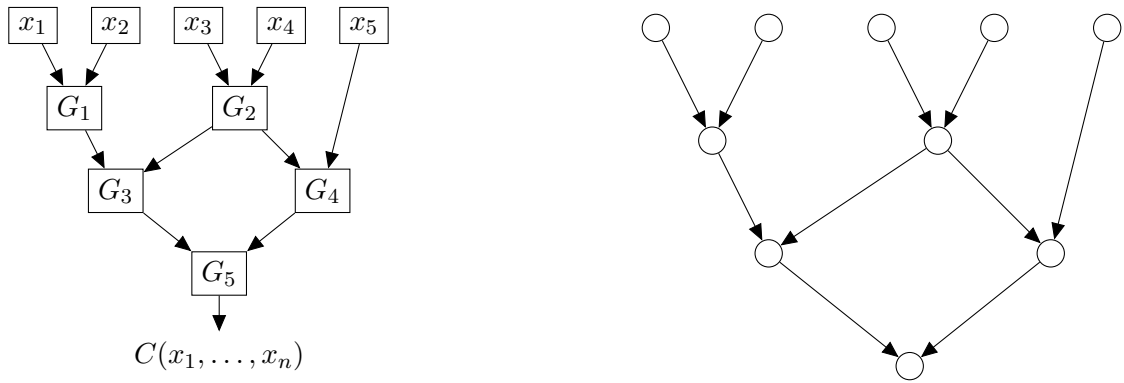

\begin{definition}[Computation by Boolean Circuits]
    A \emph{Boolean function} with $n$ inputs is a function $f : \{0,1\}^n \rightarrow \{0,1\}$. 
    
    A \emph{Boolean circuit} refers to a directed acyclic graph (DAG) where vertices with no incoming edges (\emph{source nodes}) are either \emph{inputs} or \emph{constants}, and all other vertices are \emph{logic gates}, where a logic gate is a Boolean function taking inputs from its incoming edges and sending its output to each outgoing edge. Logic gates with no outgoing edges (\emph{sink nodes}) are \emph{outputs}. See Figure \ref{fig:booleanCircuit}
    
    When working in a directed graph, by \emph{path} we always mean a \emph{directed path} - a sequence of distinct forward-facing edges joining distinct vertices sequentially. The \emph{depth} of a circuit is the length of the longest path in the DAG.
    
    A \emph{Boolean formula} is a special case of a Boolean circuit where all inputs and gates have only one outgoing edge, and there is only one output.
    
    We say a Boolean circuit with $n$ inputs $C$ \emph{computes} a Boolean function with $n$ inputs $f$ if and only if for all $(x_1,\dots,x_n) \in \{0,1\}^n$,

    \begin{equation*}
        C(x_1,\dots,x_n) \ = \ f(x_1,\dots,x_n).
    \end{equation*}

    Note that Boolean circuits can have more than one output, but if a Boolean circuit computes a Boolean function, it must have a \emph{single output}. It is a nice observation that a Boolean circuit with one output is a Boolean formula if and only if the underlying undirected graph is a tree.
\end{definition}

\begin{theorem}[Shannon, 1938 \cite{S38}. Proof omitted.]
\label{thm:shannon}
    Any Boolean function can be computed by a Boolean formula consisting of AND, OR and NOT gates, where an AND gate has two inputs and outputs 1 if and only if both inputs are 1, an OR gate has two inputs and outputs 0 if any only if both inputs are 0, and a NOT gate has one input and negates it.
    
    These are denoted $a \wedge b$, $a \vee b$, and $\neg a$ respectively.
\end{theorem}

\begin{definition}[Noisy Circuits]
\label{def:noisyCircuits}
    We define a logic gate to be \emph{$\delta$-noisy} if it has probability $\delta$ of outputting the wrong output given its inputs. Alternatively, a $\delta$-noisy gate \emph{malfunctions} with probability $\delta$. 
    
    A \emph{$\delta$-noisy circuit} is a circuit consisting of only $\delta$-noisy gates, where gates malfunction independently. 
    
    If the input or output of a gate in a noisy circuit is not equal to what it would be were the circuit noiseless, we call the input or output \emph{incorrect}.
\end{definition}

\begin{remark}
    We assume throughout that $\delta \leq \frac{1}{2}$, as if $\delta > \frac{1}{2}$, the probability of the gate acting unexpectedly remains less than $\frac{1}{2}$.
\end{remark}

In the proof of \ref{thm:shannon}, the method of Shannon's construction meant that complex functions with more inputs were computed by deeper circuits. von Neumann spotted that if Shannon's circuits were noisy, they would necessarily have a probability of incorrect computation approaching $\frac{1}{2}$ as the circuits got deeper. This led von Neumann to investigate whether circuits could be constructed in a way that avoided this.

\begin{definition}[Reliable Computation]
\label{def:rel}
    We say that a set of Boolean functions $\mathcal{F}$ can be \emph{reliably computed} by a set of $\delta$-noisy gate circuits $\mathcal{C}$ if and only if there exists $\varepsilon = \varepsilon(\delta) > 0$ s.t. for all $f \in \mathcal{F}$ with $n$ inputs, there exists $C\in \mathcal{C}$ with $n$ inputs s.t. for all $(x_1,\dots,x_n) \in \{0,1\}^n$,

    \begin{equation*}
        \Prob \left(C(x_1,\dots,x_n) \ \neq \ f(x_1,\dots,x_n)\right) \ \leq \ \frac{1}{2} \ - \ \varepsilon.
    \end{equation*}

    Also, when $\mathcal{F}$m $\mathcal{C}$, and $\varepsilon$ are defined as above, we say that $f\in\mathcal{F}$ is \emph{reliably computed} by $C\in\mathcal{C}$ if and only if for all $(x_1,\dots,x_n) \in \{0,1\}^n$,

    \begin{equation*}
        \Prob \left(C(x_1,\dots,x_n) \ \neq \ f(x_1,\dots,x_n)\right) \ \leq \ \frac{1}{2} \ - \ \varepsilon.
    \end{equation*}
\end{definition}

\begin{remark}
    Suppose Boolean function $f$ is reliably computed by noisy circuit $C$. By running the circuit $C$ with inputs $(x_1,\dots,x_n) \in \{0,1\}^n$ repeatedly, we can get an arbitrarily low probability of incorrectly computing $f(x_1,\dots,x_n)$.

    Indeed, suppose we independently compute $C(x_1,\dots,x_n)$ $m$ times, and choose to compute $f(x_1,\dots,x_n)$ as the bit which occurred $> \frac{m}{2}$ times (if we have a tie, choose randomly). Then by construction, we have

    \begin{equation}
        \Prob(\text{Compute $f$ incorrectly}) \ \leq \ \Prob\left(\text{\# errors} \ \geq \ \frac{m}{2} \right).
    \end{equation}

    The number of errors each computation of $C(x_1,\dots,x_n)$ makes is an independent Bernoulli random variable with parameter $\leq \frac{1}{2}-\varepsilon$. By Hoeffding's inequality \cite{H63}, we have

    \begin{equation}
        \Prob\left(\text{\# errors} \ \geq \ \frac{m}{2} \right) \ \leq \ \exp(-2\varepsilon^2 m).
    \end{equation}

    Thus the probability of computing $f$ incorrectly using $C$ can get arbitrarily close to 0 for large $m$.
\end{remark}

\begin{lemma}
\label{lem:threshold}
    There exists threshold $\delta^*_{\text{3MAJ}}<\frac{1}{2}$ such that reliable computation with a $\delta$-noisy circuit of 3MAJ gates is possible if \emph{and only if} $\delta \leq \delta^*_{\text{3MAJ}}$.
\end{lemma}

\begin{proof}
     We will first prove the existence of the threshold. Suppose reliable computation is possible for $\delta_0 \geq 0$ (in fact, we prove in Section \ref{sec:von} that it is possible for some sufficiently small $\delta_0 > 0$). By the definition of reliable computation, there exists $\varepsilon = \varepsilon(\delta_0) > 0$ s.t. for each $f$, there exists a $\delta_0$-noisy circuit of 3MAJ gates $C$ such that for all $(x_1,\dots,x_n) \in \{0,1\}^n$,

        \begin{equation*}
            \Prob \left(C(x_1,\dots,x_n) \ \neq \ f(x_1,\dots,x_n)\right) \ \leq \ \frac{1}{2} \ - \ \varepsilon.
        \end{equation*}
     
     We will show that reliable computation is certainly possible for $\delta \leq \delta_0$. Take the same $\varepsilon(\delta_0)$ that worked for $\delta_0$, and for each $f$ take the same circuit $C$ as above. 
     
     We aim to show that the probability that $C$ computes $f$ increases when $C$ is $\delta$-noisy compared to when $C$ is $\delta_0$-noisy. This is equivalent to showing that the probability the output of $C$ is incorrect decreases. We will proceed by induction on the length of the longest path in a circuit (or more precisely, the circuit's corresponding DAG).
     
     Consider a new circuit $C^*$ formed from removing the sink node in the corresponding DAG of circuit $C$. By the induction hypothesis, the probability the outputs of $C^*$ are incorrect decreases when $\delta$-noisy compared to when $\delta_0$-noisy. Observe that $C^*$ has multiple outputs, but we can generalize our induction hypothesis to hold for such circuits. The base case, where the longest length of a path is 1, remains trivial. Each gate has noiseless inputs, thus decreasing the noise of the gates decreases the probability that the outputs are incorrect.

     Let $P$ be the (not necessarily unique) longest path in $C$. Observe that the final gate in $P$ was removed from $C$ in the construction of $C^*$. It can easily be checked for a 3MAJ gate that if the probabilities of each input being incorrect has decreased, and if the noise of the gate decreases, it follows that the probability the output is incorrect decreases. The output of this final gate is the output of $C$, so we are done.

     
     
     Since the probability of $C$ computing $f$ correctly has increased, we have that reliable computation is possible for $\delta \leq \delta_0$. Thus there exists a largest $\delta$, namely $\delta^*_{\text{3MAJ}}$, which reliable computation is possible for.

     We will now prove the threshold must be less than $\frac{1}{2}$. Suppose $\delta=\frac{1}{2}$. By considering the final gate of the circuit leading to the output, we see that the final output of the circuit will always be 0 with probability $\frac{1}{2}$ and 1 with probability $\frac{1}{2}$. This ensures our circuit's probability of failure is $\frac{1}{2}$, when we require a failure probability less than $\frac{1}{2}$ for reliable computation. Thus  $\delta^*_{\text{3MAJ}} < \frac{1}{2}$.
\end{proof}

It remains an open problem to find the threshold $\delta^*_{\text{3MAJ}}$. Further research into reliable computation from unreliable components sought to generalize to circuits that are composed of \emph{any} gate with at most $k$ inputs.

\begin{definition}[Fan-in and Fan-out]
    The \emph{fan-in of a gate} is its number of inputs. The \emph{fan-out of a gate} is its number of inputs. 
    
    We say the \emph{fan-in of a circuit} $k$ is the maximum fan-in amongst gates in the circuit. Note that we always assume $k>1$, as fan-in 1 circuits are unable to compute any Boolean function with more than one input.
\end{definition}

Finding the threshold $\delta^*(k)$ for noisy circuits with fan-in $k$ remains open for all values of $k$. A similar method of proof as used for Lemma \ref{lem:threshold} establishes the existence of $\delta^*(k)$ for all $k$. The best known upper bound on $\delta^* (k)$ comes from the following result.

\begin{theorem}[Evans and Schulman, 1999 \cite{ES99}.]
\label{thm:es99}
    For fan-in $k$ noisy circuits, we have the threshold noise for reliable computation $\delta^* (k)$ is upper bounded by 
    
    \begin{equation*}
        \delta^* (k) \ \leq \ \delta^*_{\text{ES}} (k) \ := \ \frac{1}{2} \ - \ \frac{1}{2\sqrt{k}} 
    \end{equation*}

    for all $k$.
\end{theorem}

We will prove this result in Section \ref{sec:pol}.

The exact threshold for reliable computations is known for noisy formulas with odd fan-in. We state the result below.

\begin{theorem}[Evans and Schulman, 2003 \cite{ES03}. Proof Omitted]
\label{thm:es03}
    For odd fan-in $k$, the threshold noise for reliable computation by formulas is
    
    \begin{equation*}
        \delta^*_f (k) \ = \ \frac{1}{2} \ - \ \frac{2^{k-2}}{k \binom{k-1}{(k-1)/2}}.
    \end{equation*}
\end{theorem}

\begin{remark}
    An intuitive argument for why $k$ even is more challenging than $k$ odd is that majority or minority gates must be ``unbalanced'' when $k$ is even, i.e. the total number of input states for which the gate outputs 0 cannot be the same as number of states for which it outputs 1, due to the boundary case where there are the same number of 0s as 1s. These gates are critical to the argument of von Neumann, and as we will see in Section \ref{sec:von}, they are a natural way to control error. 

    Interestingly, the threshold remained elusive for fan-in 2, which are the most commonly used and best understood formulas, until 2007. Unger proved that $\delta^*_f (2) = \frac{3-\sqrt{7}}{4} \approx 0.08856$ \cite{U07}.

    Unger conjectured his approach can be used to establish $\delta^*_f (k)$ for $k>2$ even, and that his result even holds for fan-in 2 circuits, but these both remain elusive. We remark that for fan-in $k>2$ even, at least we know $\delta^*_f (k+1)$, and clearly $\delta^*_f (k)\leq \delta^*_f (k+1)$.
\end{remark}

We use Evans-Schulman's 2003 result about noisy formulas to establish the following lemma: 

\begin{lemma}
\label{lem:converg}
    As fan-in $k\rightarrow \infty$, we have that $\delta^* \rightarrow \frac{1}{2}$. In addition, $\delta^*$ and $\delta^*_{\text{ES}}$ have the same rate of convergence to $\frac{1}{2}$.
\end{lemma}

\begin{remark}
    It is a rather remarkable result that when fan-in is large, even a Boolean function with $n \gg k$ inputs can be reliably computed by a circuit consisting of gates that are not much better than a coin flip. This lemma also allows us to appreciate that the Evans-Schulman upper bound is asymptotically tight to the true value of $\delta^*$.
\end{remark}

\begin{figure}
    \includegraphics[scale=0.2]{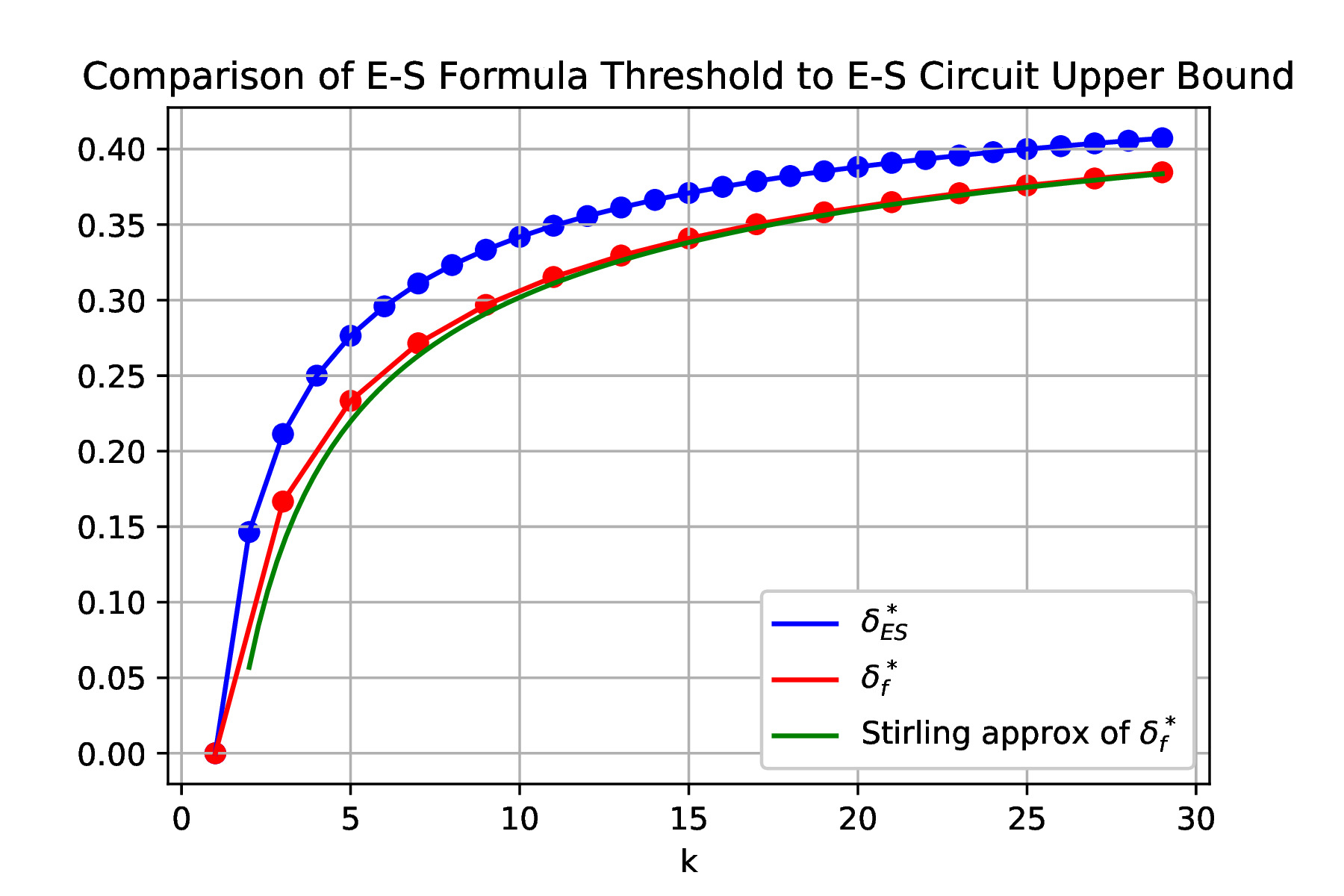}
    \label{fig:comparison}
    \caption{Scatter points mark the values at positive integers and odd positive integers for $\delta^*_{\text{ES}}$ and $\delta^*_f$ respectively.}
\end{figure}

\begin{proof}
    Since formulas are a type of circuit, we have $\delta^*_f \leq \delta^* \leq \delta^*_{\text{ES}}$ by Theorems \ref{thm:es99} and \ref{thm:es03}. For odd fan-in, this implies that $\delta^*$ must lie between the red and blue points on the odd integers in Figure \ref{fig:comparison}. Since $\delta^*_f \rightarrow \frac{1}{2}$ as $k \rightarrow \infty$, we also have $\delta^* \rightarrow \frac{1}{2}$ as $k\rightarrow \infty$.

    By Stirling's approximation, we have

    \begin{equation}
        \delta^*_f (k) \ \approx \ \frac{1}{2} \ - \ \frac{\sqrt{\pi}}{2\sqrt{2k}}.
    \end{equation}
    
    This is an asymptotically equal approximation for $k \rightarrow \infty$. Thus $\delta^*_f$ and $\delta^*_{\text{ES}}$ have the same rate of convergence, and as they sandwich $\delta^*$, this implies that $\delta^*$ and $\delta^*_{\text{ES}}$ have the same rate of convergence.
\end{proof}

\section{Reliable Computation from Unreliable 3-input Majority/Minority Gates}
\label{sec:von}

In this section, we will prove the first major result in the field of reliable computation.

\begin{theorem}[von Neumann, 1952]
\label{thm:von}
    Provided $\delta$ is sufficiently small, any Boolean function that can be computed by a circuit of 3-input majority (3MAJ) gates can be reliably computed by a $\delta$-noisy circuit of 3MAJ gates.
\end{theorem}

This result implies that $\delta^*_{\text{3MAJ}}>0$. We will also prove a slightly different result.

\begin{theorem}
\label{thm:3MIN}
    Provided $\delta$ is sufficiently small, any Boolean function can be reliably computed by a $\delta$-noisy circuit composed of 3-input \emph{minority} gates.
\end{theorem}

In some sense, this is a stronger result, as we are now able to reliably compute all Boolean functions. However, the method of proof is almost exactly the same as von Neumann's.

The truth table for a 3MAJ gate is given in Table \ref{tab:maj}, and the truth table for a 3-input minority (3MIN) gate is given in Table \ref{tab:notmaj}.

\begin{table}
\centering
\begin{minipage}{0.45\textwidth}
\centering
\caption{Truth table for a 3MAJ gate}
\label{tab:maj}
\begin{tabular}{|c|c|c||c|}
\hline
$a$ & $b$ & $c$ & Output \\
\hline
0 & 0 & 0 & 0 \\
0 & 0 & 1 & 0 \\
0 & 1 & 0 & 0 \\
0 & 1 & 1 & 1 \\
1 & 0 & 0 & 0 \\
1 & 0 & 1 & 1 \\
1 & 1 & 0 & 1 \\
1 & 1 & 1 & 1 \\
\hline
\end{tabular}
\end{minipage}
\begin{minipage}{0.45\textwidth}
\centering
\caption{Truth table for a 3MIN gate}
\label{tab:notmaj}
\begin{tabular}{|c|c|c||c|}
\hline
$a$ & $b$ & $c$ & Output \\
\hline
0 & 0 & 0 & 1 \\
0 & 0 & 1 & 1 \\
0 & 1 & 0 & 1 \\
0 & 1 & 1 & 0 \\
1 & 0 & 0 & 1 \\
1 & 0 & 1 & 0 \\
1 & 1 & 0 & 0 \\
1 & 1 & 1 & 0 \\
\hline
\end{tabular}
\end{minipage}
\end{table}

\begin{lemma}
    A circuit of 3MAJ gates is unable to compute the Boolean function NOT.
\end{lemma}

\begin{proof}
    Let us call the input of our circuit $a \in \{0,1\}$. The three possible inputs for a 3MAJ gate at the beginning of the computation are $a$, constant 0, and constant 1. We break down the outcomes of different inputs into the following three cases:

    \begin{itemize}
        \item Only constant inputs result in a constant output of 0 or 1.
    
        \item One $a$ input, one $0$ input, and one $1$ input results in an output of $a$.

        \item One $a$ input and either two constant 0 inputs or two constant 1 inputs results in a constant output of 0 or 1 respectively.

        \item Two or three $a$ inputs results in an output of $a$.
    \end{itemize}

    Thus the only possible outputs of a 3MAJ gate circuit with input $a$ are $a$, constant 0, or constant 1.
\end{proof}

\begin{lemma}
\label{lem:complete}
    A formula of 3MIN gates is able to compute any Boolean function, or in other words, is \emph{functionally complete}.
\end{lemma}

\begin{proof}
    We just need to prove that 3MIN gates can simulate AND, OR, and NOT gates, as then we are done by Theorem \ref{thm:shannon}. We can easily make a NOT gate from a 3MIN gate. Take one input to be the variable $a$, one input to constantly be 0, and one input to constantly be 1. This outputs $\neg a$.

    A NAND gate is a composition of an AND gate and then a NOT gate - it has two inputs, and outputs 0 if and only if both inputs are 1. We can also make a NAND gate from a 3MIN gate. Take one input to be a variable $a$, one input to be a variable $b$, and one input to constantly be 0. This outputs $\neg (a\wedge b)$.

    Similarly, a NOR gate is a composition of an OR gate and then a NOT gate - it has two inputs, and outputs 1 if and only if both inputs are 0. To make a NOR gate from a 3MIN gate, take one input to be a variable $a$, one input to be a variable $b$, and one input to constantly be 1. This outputs $\neg (a\vee b)$.

    We can now produce an AND gate from 3MIN by composing our NAND gate and then our NOT gate, and also produce an OR gate by composing our NOR gate and then our NOT gate. So we are done.
\end{proof}

For the remainder of this section, if we do not specify what type of gate we are discussing, we refer to either a 3MIN or 3MAJ gate. We now give two important lemmas about such gates.

\begin{lemma}
\label{lem:generalBound}
    For some fixed inputs and $\delta$-noisy circuit $C$, consider one of its gates, $G$. Let $0 < \eta_i \leq 1$, $i=1,2,3$, be an upper bound for the probability that the $i^{\text{th}}$ input into $G$ is incorrect. Then

    \begin{equation*}
        \delta \ + \ \eta_1 \ + \ \eta_2 \ + \ \eta_3
    \end{equation*}

    is an upper bound for the probability that the output of $G$ is incorrect.
\end{lemma}

\begin{proof}
    Let $A_i$ be the event that input $i$ is incorrect (so $\Prob(A_i) \leq \eta_i$) and let $B$ be the event that $G$ malfunctions. We have

    \begin{equation}
        \Prob(\text{Output of $G$ is incorrect}) \ \leq \ \Prob\left( \bigcup_i A_i \ \cup \ B \right) \ \leq \ \delta \ + \ \eta_1 \ + \ \eta_2 \ + \ \eta_3
    \end{equation}

    as required, where the second inequality is due to Boole's inequality.
\end{proof}

\begin{lemma}
\label{lem:specialBound}
    Assume the same definitions and conditions as Lemma \ref{lem:generalBound}, and also define $A_i$ and $B$ in the same way. Further assume that the events $A_i$ are independent, and also that if the inputs are correct, all three should be equal. Then

    \begin{equation}
        \Theta \ := \ \eta_1\eta_2 \ + \ \eta_1\eta_3 \ + \ \eta_2\eta_3 \ - \ 2\eta_1\eta_2\eta_3
    \end{equation}

    is an upper bound probability for at least two of the input lines of $G$ being incorrect. Thus

    \begin{equation}
        \delta^\prime \ := \ (1-\delta)\Theta \ + \ \delta(1-\Theta) \ = \ \delta \ + \ (1-2\delta)\Theta
    \end{equation}

    is a \emph{smaller} upper bound for the probability that the output of $G$ is incorrect than Lemma \ref{lem:generalBound}.
\end{lemma}

\begin{proof}
    We can show $\Theta$ is an upper bound probability for at least two of the input lines of $G$ being incorrect by inclusion-exclusion. It follows immediately that $\delta^\prime$ is an upper bound for the probability that the output of $G$ is incorrect.

    We have that

    \begin{equation}
        \delta^\prime \ = \ \delta \ + \ (1-2\delta)\Theta \ < \ \delta \ + \ \Theta.
    \end{equation}

    By the definition of $\Theta$ and Boole's inequality, we have

    \begin{equation}
        \Theta \ \leq  \Prob\left( \bigcup_i A_i \right) \ \leq \ \eta_1 \ + \ \eta_2 \ + \ \eta_3,
    \end{equation}

    thus indeed $\delta^\prime < \delta + \eta_1 + \eta_2 + \eta_3$ and we are done.
\end{proof}

We will now prove Theorem \ref{thm:3MIN}, i.e. that provided $\delta$ is sufficiently small, any Boolean function can be reliably computed by a $\delta$-noisy circuit composed of 3MIN gates.

\begin{figure}
    \includegraphics[scale=0.30]{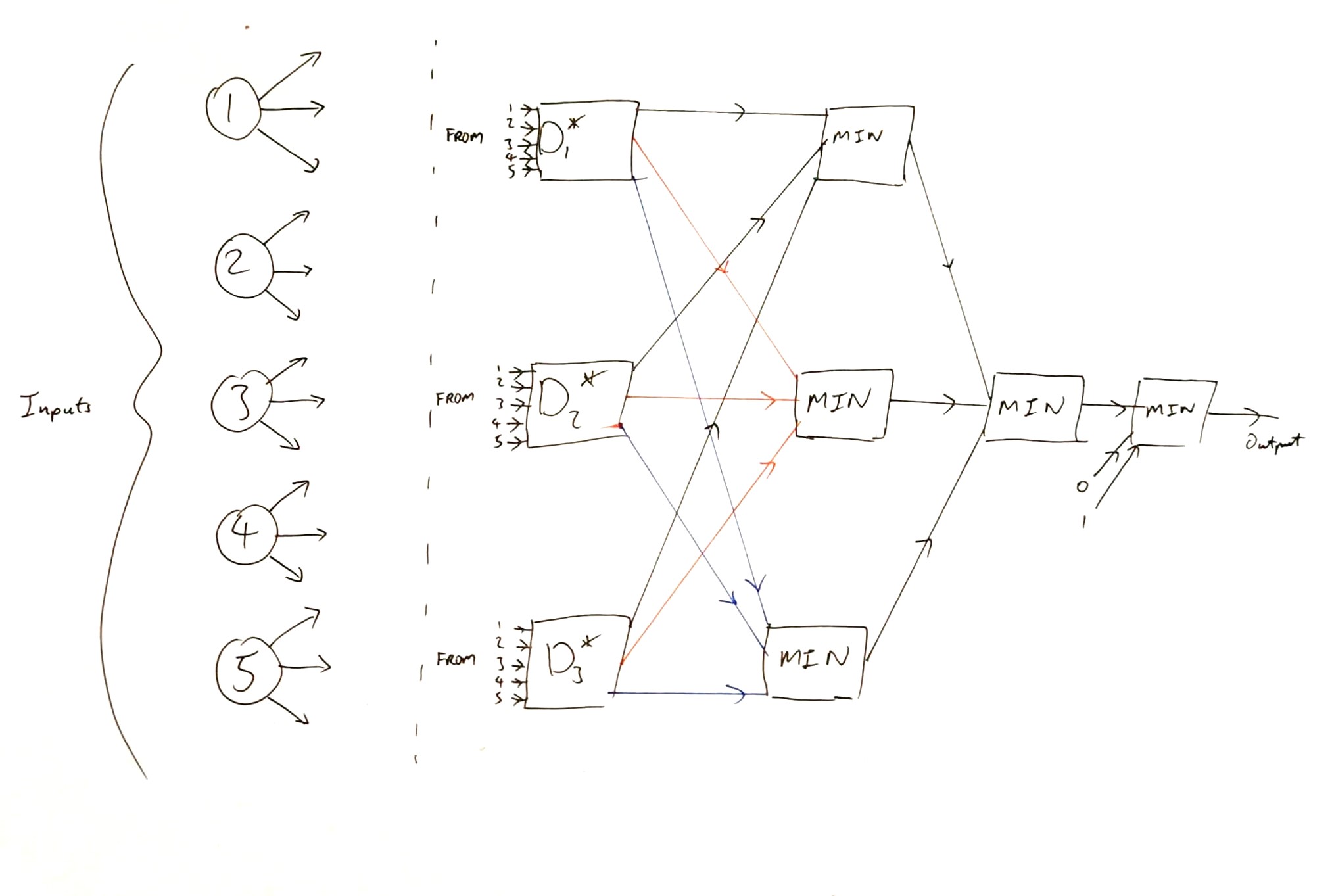}
    \caption{Construction of $C^*$ in the case $n=5$. This is the first level of the induction, where $C$ has one output and computes $f$.}
    \label{fig:construction}
\end{figure}

\begin{figure}
    \includegraphics[scale=0.39]{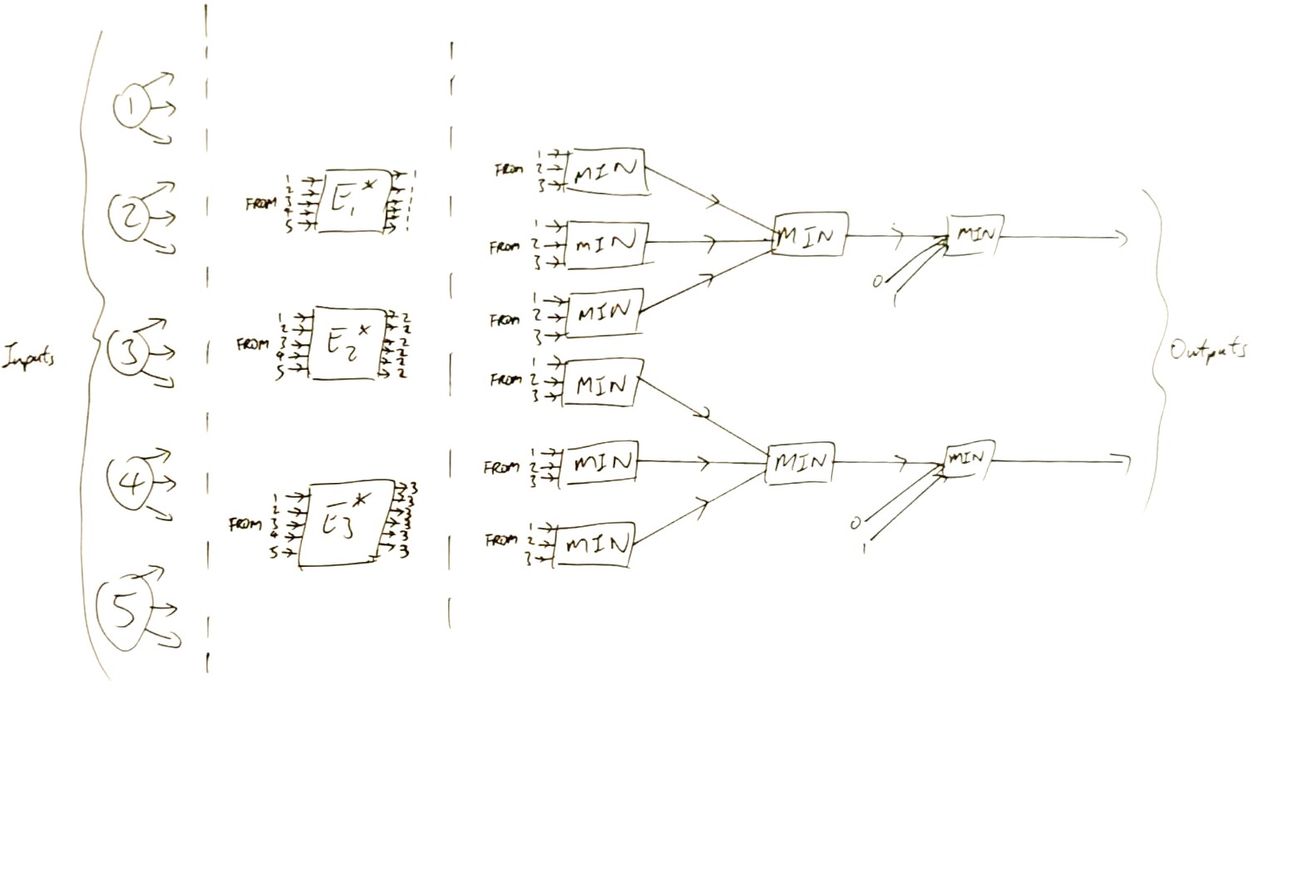}
    \caption{Construction of $D^*$ in the case $n=5$ and $D$ has 2 outputs ($m=2$).}
    \label{fig:harderConstruction}
\end{figure}

\begin{figure}
    \includegraphics[scale=0.2]{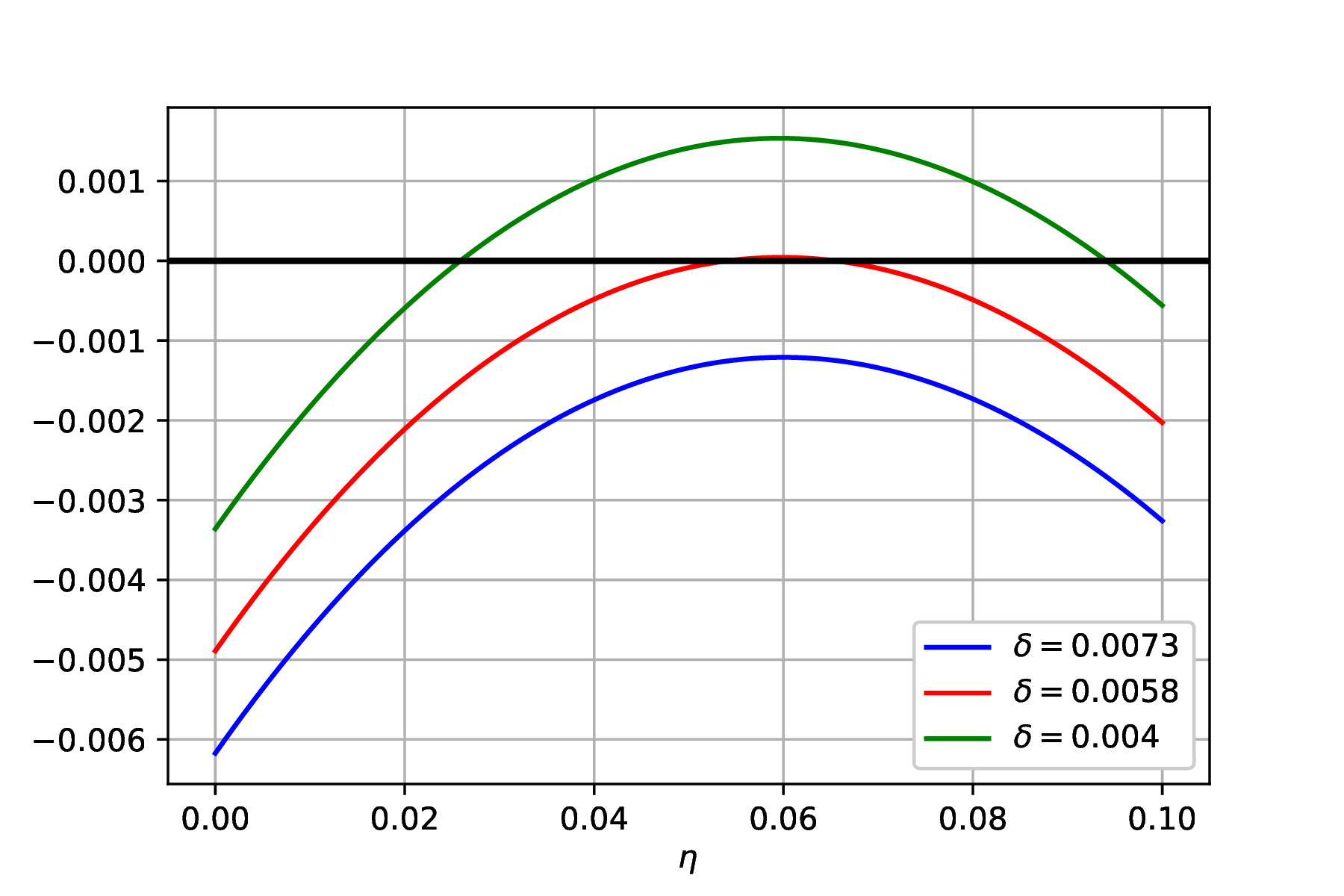}
    \caption{Plot of cubic polynomial $p_\delta (\eta)$ in the 3MIN gates proof for three different values of $\delta$.}
    \label{fig:minproof}
\end{figure}

\begin{proof}[Proof of Theorem \ref{thm:3MIN}]
    We want to show that there exists sufficiently small $\delta$ s.t. there exists $\varepsilon = \varepsilon(\delta) > 0$ s.t. for all Boolean functions $f$, there exists $\delta$-noisy circuit composed of 3MIN gates $C$ s.t. for all $(x_1,\dots,x_n) \in \{0,1\}^n$,

    \begin{equation*}
        \Prob \left(C(x_1,\dots,x_n) \ \neq \ f(x_1,\dots,x_n)\right) \ \leq \ \frac{1}{2} \ - \ \varepsilon.
    \end{equation*}

    Let $\eta(\delta) := \frac{1}{2} - \varepsilon(\delta)$. For each Boolean function $f$, we will inductively construct a $\delta$-noisy circuit composed of 3MIN gates that computes $f$ with probability of error upper bounded by $\eta$.

    Let $C$ be a circuit consisting of 3MIN gates that computes $f$ in the normal, noiseless setting. This certainly exists by Lemma \ref{lem:complete}. Let $P$ be the longest path in $C$ from an input to the output. If $P$ has length zero, there is nothing to prove, so we assume it is not.

    Construct circuit $D$ by removing the output gate from $C$. Note that the circuit $D$ has three output gates. By induction on the length of the longest path, which we have now decreased by 1, there exists a $\delta$-noisy circuit composed of 3MIN gates $D^*$ that is equivalent to $D$, but each output is incorrect with probability upper bounded by $\eta$.

    Produce three identical copies of $D^*$, namely $D^*_1,D^*_2,D^*_3,$. We now aim to produce a circuit $C^*$ that simulates $C$ with an upper bound $\eta$ on the probability of error in its output. We do this by sending the first output of each of $D^*_1,D^*_2,D^*_3,$ to a 3MIN gate, and similar for the second and third output. We then send the outputs of these three gates to another 3MIN gate, and finally invert the output with a NOT gate made from a 3MIN gate, as described in the proof of Lemma \ref{lem:complete}. 
    
    The inclusion of the NOT gate ensures that the correct output of $C^*$ matches what $C$ outputs. In $C$, the outputs of $D$ go through another 3MIN gate. But in $C^*$, we have sent the outputs of $D^*$ through two 3MIN gates in series, leading to an inversion we must correct. See Figure \ref{fig:construction}.
    
    Note that this inductive construction requires more care when if we wish to construct $D^*$ simulating $D$ when $D$ has multiple outputs. Producing $C^*$ is an easy case because $C$ only has a single output. When $D$ has $m$ outputs, let $E$ be the circuit produced by removing all of $D$'s output gates. $E$ has at most $3m$ outputs. If it does not attain this, it is due to outputs of $E$ leading to multiple outputs of $D$. We can w.l.o.g. assume $E$ has $3m$ outputs, as if some output gate $G \in E$ led to $l>1$ of $D$'s outputs, we can replicate $G$ until there are $l$ of them. After doing this for all such outputs of $E$, $E$ now has $3m$ outputs, and after connecting the new outputs of $E$ to the appropriate output of $D$, the behaviour of $D$ is unaffected. Note that this replication process does not change the length of the longest path in $E$.
    
    We have decreased the length of the longest path by 1, so there exists noisy circuit $E^*$ that simulates $E$ with each of its outputs having probability of error upper bounded by $\eta$ by induction. Triplicate $E^*$ as before, and send the $3m$ outputs to $3m$ 3MIN gates as in Figure \ref{fig:harderConstruction}. Then send these outputs to $m$ 3MIN gates, and invert them, to construct $D^*$.

    This inductive process eventually yields a full construction of $C^*$. We are interested in when $C^*$'s output's probability of error is smaller than $\eta$, i.e. if we have
    
    \begin{equation*}
        \Prob \left(C^* (x_1,\dots,x_n) \ \neq \ f(x_1,\dots,x_n)\right) \ \leq \ \eta \ = \ \frac{1}{2} \ - \ \varepsilon,
    \end{equation*}    
    
    we would be done, as in each step of our induction, the error of any output never exceeds $\eta$. By Lemma \ref{lem:specialBound}, the first round of 3MIN gates in our construction are incorrect with probability upper bounded by
    
    \begin{equation}
    \label{eqn:layer1}
        g_\delta (\eta) \ := \ \delta \ + \ (1-2\delta)(3\eta^2 - 2\eta^3). 
    \end{equation}

    By Lemma \ref{lem:generalBound} and (\ref{eqn:layer1}), the next round of 3MIN gates are incorrect with probability upper bounded by

    \begin{equation}
    \label{eqn:layer2}
        \delta \ + \ 3g_\delta (\eta).
    \end{equation}

    Again by Lemma \ref{lem:generalBound} and now (\ref{eqn:layer2}), the output of the final inversion is incorrect with probability upper bounded by

    \begin{equation}
        2\delta \ + \ 3g_\delta (\eta),
    \end{equation}

    noting that the constant inputs in this gate are always correct.

    This means we require the condition

    \begin{equation}
        2\delta \ + \ 3g_\delta (\eta) \ \leq \ \eta,
    \end{equation}

    i.e.

    \begin{equation}
        \eta^3 \ - \ \frac{3}{2} \eta^2 \ + \ \frac{1}{6(1-2\delta)}\eta \ - \ \frac{5\delta}{6(1-2\delta)} \ \geq \ 0.
    \end{equation}

    Indeed, when $\delta < 0.0058$,

    \begin{equation}
        p_\delta (\eta) \ := \ \eta^3 \ - \ \frac{3}{2} \eta^2 \ + \ \frac{1}{6(1-2\delta)}\eta \ - \ \frac{5\delta}{6(1-2\delta)} \ = \ 0.
    \end{equation}

    has a root $\eta \in \left(0,\frac{1}{2}\right)$. See Figure \ref{fig:minproof} for an illustration - we can see that there exist $\eta$ satisfying $p_\delta (\eta) \geq 0$ when $\delta = 0.004$ and 0.0058, but not when $\delta = 0.0073$. It can easily be shown that the condition holds for all $\delta < 0.0058$. Thus for $\delta$ sufficiently small, we have reliable computation using a noisy circuit with 3MIN gates. 
\end{proof}

We will now prove Theorem \ref{thm:von}, i.e. that provided $\delta$ is sufficiently small, any Boolean function that can be computed by a circuit of 3MAJ gates can be reliably computed by a $\delta$-noisy circuit composed of 3MAJ gates.

\begin{figure}
    \includegraphics[scale=0.2]{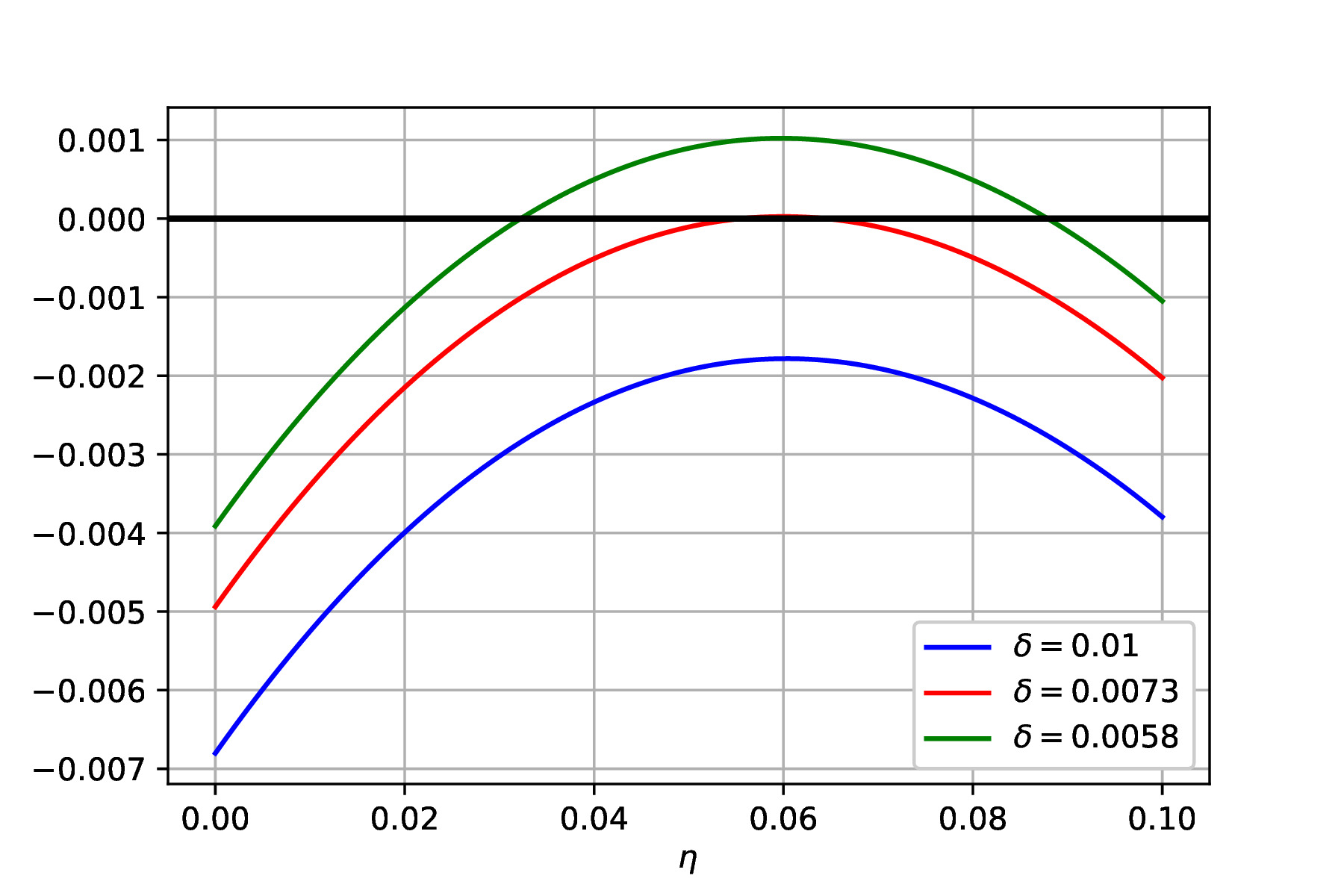}
    \caption{Plot of cubic polynomial $q_\delta (\eta)$ in the 3MAJ gates proof for three different values of $\delta$.}
    \label{fig:majproof}
\end{figure}

\begin{proof}[Proof of Theorem \ref{thm:von}]
    Essentially the same argument as in the proof of Theorem \ref{thm:3MIN}. This time $C$ is a circuit of 3MAJ gates that computes $f$. The inductive construction of $C^*$ also no longer needs a NOT gate at the end, as the extra 3MAJ gate does not cause an inversion. This means that we only require the slightly looser condition

    \begin{equation}
        \delta \ + \ 3g_\delta (\eta),
    \end{equation}

    i.e.

    \begin{equation}
        \eta^3 \ - \ \frac{3}{2} \eta^2 \ + \ \frac{1}{6(1-2\delta)}\eta \ - \ \frac{2\delta}{3(1-2\delta)} \ \geq \ 0.
    \end{equation}

    Indeed, when $\delta < 0.0073$,

    \begin{equation}
        q_\delta (\eta) \ := \ \eta^3 \ - \ \frac{3}{2} \eta^2 \ + \ \frac{1}{6(1-2\delta)}\eta \ - \ \frac{2\delta}{3(1-2\delta)} \ = \ 0.
    \end{equation}

    has a root $\eta \in \left(0,\frac{1}{2}\right)$. See Figure \ref{fig:majproof} for an illustration - we can see that there exist $\eta$ satisfying $q_\delta (\eta) \geq 0$ when $\delta = 0.0058, 0.0073$, but not $\delta = 0.01$. It can easily be shown that the condition holds for all $\delta < 0.0073$. Thus for $\delta$ sufficiently small, we have reliable computation using a noisy circuit with 3MAJ gates. 
\end{proof}

We believe that this extension to von Neumann's work could be significant. Unger's 2007 work finding the threshold $\delta^*_f (2)$ for fan-in 2 formulas built on an earlier paper by Evans and Pippenger in 1998 that found the threshold for formulas of 2-input NAND gates $\delta^*_{\text{NAND},f}$ \cite{EP98}. Since NAND is functionally complete, we immediately have $\delta^*_{\text{NAND},f} \leq \delta^*_f (2)$. Remarkably, the thresholds are both equal to $\frac{3-\sqrt{7}}{4}$.

This suggests that for formulas or circuits with some fan-in $k$, a restriction onto just a single functionally complete fan-in $k$ gate may not affect the threshold for noisy computation. In light of this result, we may make the following conjecture:

\begin{conjecture}
    Let $\delta^*_{\text{3MIN}}$ be the threshold for reliable computation with noisy circuits of 3MIN gates. Then
    
    \begin{equation}
        \delta^*_{\text{3MIN}} \ = \ \delta^* (3).
    \end{equation}.
\end{conjecture}

We certainly have  $\delta^*_{\text{3MIN}} \leq \delta^* (3)$ since 3MIN is functionally complete, and determining $\delta^*_{\text{3MIN}}$ may be easier than determining $\delta^* (3)$.


\section{Strong Data-Processing Inequalities}

In this section we will introduce the necessary information theoretic techniques required to define the Strong Data-Processing Inequality used to prove Theorem \ref{thm:es99}. We restrict our attention only to what is needed. For all omitted proofs, a more extensive overview, and measure-theoretic generalisation of the following techniques, see \cite{PWTB}.

For the remainder of the essay, all logarithms are assumed to be base 2, unless stated otherwise. Also observe that we take the value of $0\log 0$ to be 0, as this is the value of the right limit.

\begin{definition}[Entropy and Conditional Entropy]
    For discrete random variable $X$ taking values in $\mathcal{X}$ with distribution $P_X$, we define the \emph{entropy} of $X$ to be

    \begin{equation*}
        H(X) \ := \ - \ \sum_{x\in\mathcal{X}} P_X(x) \log P_X(x).
    \end{equation*}

    We can also define the entropy of a distribution by $H(P_X) = H(X)$, where $X\sim P_X$. For $X\sim$ Bernoulli($\alpha$), we often write

    \begin{equation*}
        H(X) \ = \ H(\alpha) \ := \ -\alpha \log \alpha \ - \  (1-\alpha) \log (1-\alpha).
    \end{equation*}

    Let $Y\sim P_Y$ and take values in $\mathcal{Y}$. We also define the \emph{conditional entropy} of $X|Y$ by

    \begin{equation*}
        H(X|Y) \ := \ \sum_{y\in\mathcal{Y}} P_Y (y) H(P_{X|Y=y}).
    \end{equation*}
\end{definition}

\begin{definition}[Relative Entropy or KL-divergence]
    For discrete probability distributions $P$ and $Q$ defined on the same sample space $\mathcal{X}$, the \emph{relative entropy} or \emph{KL-divergence} of $P$ from $Q$ is defined to be

    \begin{equation*}
        D(P\parallel Q) \ := \ \sum_{x\in\mathcal{X}} P(x) \log \frac{P(x)}{Q(x)}.
    \end{equation*}

    For continuous distributions $P$ and $Q$ with probability densities $p$ and $q$ respectively, \emph{relative entropy} is defined to be the integral

    \begin{equation*}
        D(P\parallel Q) \ := \ \int_{-\infty}^\infty p(x) \log \frac{p(x)}{q(x)}\text{d}x.
    \end{equation*}
\end{definition}

\begin{definition}[Mutual Information and Conditional Mutual Information]
    Let $(X,Y)$ be a pair of random variables with values over the space $\mathcal{X} \times \mathcal{Y}$. Denote their joint distribution $P_{X,Y}$ and marginal distributions $P_X$ and $P_Y$ respectively. The \emph{mutual information} between $X$ and $Y$ is defined as

    \begin{equation*}
        I(X;Y) \ := \ D(P_{X,Y} \parallel P_X P_Y).
    \end{equation*}

    We also define the \emph{conditional mutual information} by

    \begin{equation*}
        I(X;Y|Z) \ := \ \Exp_Z D(P_{X,Y |Z} \parallel P_{X|Z} P_{Y|Z}).
    \end{equation*}    
\end{definition}

\begin{remark}
    We can think of entropy $H(X)$ as the average level of uncertainty inherent to the $X$'s possible outcomes, or in other words, the amount of \emph{information} gained by observing the r.v. $X$. We can think of conditional entropy $H(X|Y)$ as the level of uncertainly inherent to $X$ given the value of $Y$ is known.

    We can think of the relative entropy as a measure of the distance of the distributions $P$ from $Q$. It is non-negative and equal to 0 if and only if the distributions are equal. It is also unbounded - an easy example of this is taking $P$ to be Bernoulli$\left(\frac{1}{2}\right)$, and $Q$ to be Bernoulli($q$). We have
    
    \begin{equation}
        D(P\parallel Q) \ = \ \frac{1}{2}(-2\log 2 \ - \ \log q \ - \ \log (1-q)).
    \end{equation}
    
    Note that $-\log q - \log (1-q)$ is unbounded for both $q\to0$ and $q\to 1$.

    Loosely, the mutual information quantifies the amount of information learned about one variable by observing the other. Unlike relative entropy, it is symmetric. Additionally, it is non-negative, and equal to 0 if and only if $X$ and $Y$ are independent. It is upper bounded by the Shannon entropy of $X$ and the Shannon entropy of $Y$. This is attained if and only if $X=Y$, or in other words,

    \begin{equation}
    \label{eqn:entropyAndMutualInfo}
        H(X) \ = \ I(X;X).
    \end{equation}

    We also have the following important relationship between conditional entropy and mutual information:

    \begin{equation}
    \label{eqn:condEntropyAndMutualInfo}
        H(Y) \ - \ H(Y|X) \ = \ H(X) \ - \ H(X|Y) \ = \ I(X;Y).
    \end{equation}

    Another standard result we will use is the chain rule for entropy and mutual information.

    \begin{equation}
    \label{eqn:chainrule}
        H(X,Y) \ = \ H(X) \ + \ H(Y|X), \ \ \ I(X;Y,Z) \ = \ I(X;Z) \ + \ I(X;Y|Z).
    \end{equation}
\end{remark}

We are now a position to introduce the Data-Processing Inequality.

\begin{lemma}[Data-Processing Inequality (DPI) for Relative Entropy, proof omitted. See Theorem 2.17 in \cite{PWTB}.]
\label{lem:dpiRelativeEntropy}
    For fixed conditional distribution $P_{Y|X}$, suppose we have prior distributions $P_X$ and $Q_X$ on the same sample space. Define posterior distributions $P_Y = P_{Y|X}\circ P_X$ and $Q_Y = P_{Y|X}\circ Q_X$. We then have

    \begin{equation*}
        D(P_X \parallel Q_X) \ \geq \ D(P_Y \parallel Q_Y).
    \end{equation*}
\end{lemma}

\begin{definition}[Contraction Coefficient for $P_{Y|X}$]
\label{def:contraction}
    For a fixed conditional distribution $P_{Y|X}$, define the \emph{contraction coefficient}
    
    \begin{equation*}
        \eta\left(P_{Y|X}\right) \ := \ \sup_{P_X,Q_X} \frac{D(P_Y \parallel Q_Y)}{D(P_X \parallel Q_X)}
    \end{equation*}

    where $P_Y = P_{Y|X}\circ P_X$ and $Q_Y = P_{Y|X}\circ Q_X$, and the supremum is taken over all pairs $(P_X,Q_X)$ satisfying $0 < D(P_X \parallel Q_X) <\infty$.
\end{definition}

\begin{corollary}[Strong Data-Processing Inequality (SDPI) for Relative Entropy]
\label{cor:sdpiRelativeEntropy}
    For fixed conditional distribution $P_{Y|X}$, and distributions $P_X$, $Q_X$, $P_Y = P_{Y|X}\circ P_X$ and $Q_Y = P_{Y|X}\circ Q_X$ defined as in Theorem \ref{lem:dpiRelativeEntropy}, we have

    \begin{equation*}
        \eta\left(P_{Y|X}\right) D(P_X \parallel Q_X) \ \geq \ D(P_Y \parallel Q_Y).
    \end{equation*}
\end{corollary}

\begin{proof}
    This is trivial from Definition \ref{def:contraction}.
\end{proof}

We also have equivalent formulations of the DPI and SDPI using mutual information. We will use the notation $U\rightarrow X \rightarrow Y$ to denote a Markov chain, i.e. $Y$ is independent of $U$ conditional on $X$.

\begin{lemma}[DPI for Mutual Information, proof omitted. See Theorem 3.7 (c) in \cite{PWTB}.]
\label{lem:dpiMutualInformation}
    For a Markov Chain $U\rightarrow X \rightarrow Y$, we have

    \begin{equation*}
        I(U;X) \ \geq \ I(U;Y).
    \end{equation*}
\end{lemma}

\begin{theorem}[SDPI for Mutual Information, full proof omitted.]
\label{thm:sdpiMutualInformation}
    For a Markov Chain $U\rightarrow X \rightarrow Y$, let $P_{U,X}$ denote the joint distribution of $U$ and $X$. Let $\eta\left(P_{Y|X}\right)$ be defined as in Definition \ref{def:contraction}. We have the following three results:

    \begin{align*}
        \text{(a)} & \ \ \ \ \eta\left(P_{Y|X}\right) \ = \ \sup_{P_{U,X}} \frac{I(U;Y)}{I(U;X)}, \\
        \text{(b)} & \ \ \ \ \eta\left(P_{Y|X}\right) I(U;X) \ \geq \ I(U;Y), \\
        \text{(c)} & \ \ \ \ \eta\left(P_{Y|X}\right) H(X) \ \geq \ I(X;Y).
    \end{align*}
\end{theorem}

\begin{proof}[Sketch of proof of Theorem \ref{thm:sdpiMutualInformation}]
    (b) with (\ref{eqn:entropyAndMutualInfo}) implies (c). We give a proof of (b) and a sketch of a proof of (a), illustrating the equivalence of the relative entropy SDPI and the mutual information SDPI.

    Denote $\eta = \eta\left(P_{Y|X}\right)$. Observe that for any $u_0$, we have

    \begin{equation}
        \eta D(P_{X|U=u_0} \parallel P_X) \ \geq \ D(P_{Y|U=u_0} \parallel P_Y)
    \end{equation}

    by definition of $\eta$. For any distribution $P_U$ of $U$, we have 

    \begin{align}
    \begin{split}
        \Exp_{P_U} \left[ \eta D(P_{X|U} \parallel P_X) \right] \ &\geq \ \Exp_{P_U} \left[ D(P_{Y|U} \parallel P_Y) \right] \\
        \implies \eta D(P_{X,U} \parallel P_X P_U) \ &\geq \ D(P_{Y,U} \parallel P_Y P_U) \\
        \implies \eta I(U;X) \ &\geq \ I(U;Y),
    \end{split}
    \end{align}

    proving (b). It remains to show the equality under supremum to yield (a). We give a brief sketch of the proof that the equality can be attained when $U\sim$ Bernoulli($\lambda$).

    Define $P_{X|U}$ by $P_{X|U=0} = \tilde{P}_X$ and $P_{X|U=1} = \tilde{Q}_X$. It can be shown that

    \begin{equation}
        I(U;X) \ = \ \lambda D(\tilde{P}_X\parallel \tilde{Q}_X) \ + \ o(\lambda)
    \end{equation}

    and

    \begin{equation}
        I(U;Y) \ = \ \lambda D(\tilde{P}_Y\parallel \tilde{Q}_Y) \ + \ o(\lambda)
    \end{equation}

    as $\lambda \to 0$ (see Appendix A.3 in \cite{PW17}). Thus

    \begin{equation}
        \frac{I(U;Y)}{I(U;X)} \ \to \ \frac{D(\tilde{P}_Y\parallel \tilde{Q}_Y)}{D(\tilde{P}_X\parallel \tilde{Q}_X)}. 
    \end{equation}

    Thus by Definition \ref{def:contraction} of $\eta$, we can take the supremum over $\tilde{P}_X$ and $\tilde{Q}_X$ and get $\eta$ arbitrarily close to $\frac{I(U;Y)}{I(U;X)}$, proving the result (a). 
\end{proof}

\begin{remark}
    Theorem \ref{thm:sdpiMutualInformation} shows that for fixed channel $P_{Y|X}$, the SDPI for mutual information and SDPI for relative entropy share the same contraction coefficient. The mutual information results are more intuitive to understand. The DPI states that there is ``some'' loss of information of how much we know about $U$ after passing $X$ through the noisy channel $P_{Y|X}$. The contraction coefficient and SDPI quantify how much guaranteed loss in information there is, regardless of the distributions of $U$ or $X$.
\end{remark}

\begin{definition}[Binary Symmetric Channel]
    Define the \emph{$\delta$-noisy binary symmetric channel} ($\text{BSC}_\delta)$) to be the conditional distribution $P_{Y|X}$ where $X$ takes values 0 and 1, and

    \begin{equation*}
        \Prob(Y=1|X=1) \ = \ \Prob(Y=0 |X=0) \ = \ 1-\delta, \ \Prob(Y=1|X=0) \ = \ \Prob(Y=0 |X=1) \ = \ \delta.
    \end{equation*}
\end{definition}

We will show that the contraction coefficient for the BSC is $\eta(\text{BSC}_\delta) = (1-2\delta)^2$. This gives us the SDPI used in the proof of Theorem \ref{thm:es99} in Section \ref{sec:pol}. We first need the following definition and result:

\begin{definition}[Squared Hellinger Distance]
    For discrete distributions $P$ and $Q$ on sample space $\mathcal{X}$, we define the \emph{Squared Hellinger Distance}

    \begin{equation*}
        H^2(P,Q) \ = \ \sum_{x\in\mathcal{X}} \left( \sqrt{P(x)} - \sqrt{Q(x)} \right)^2.
    \end{equation*}
\end{definition}

\begin{theorem}[Squared Hellinger Distance bounds on Contraction Coefficient, proof omitted. See Theorem 33.6 (e) in \cite{PWTB}.]
\label{thm:hellingerbound}
    For binary-input channels $P_{Y|X}$, denote $P_0 = P_{Y|X=0}$ and $P_1 = P_{Y|X=1}$. Then

    \begin{equation*}
        \frac{1}{2} H^2(P_0,P_1) \ \leq \ \eta\left(P_{Y|X}\right) \ \leq \ H^2(P_0,P_1) \ - \ \frac{(H^2(P_0,P_1))^2}{4}
    \end{equation*}
\end{theorem}

\begin{lemma}[Upper bound on $\eta(\text{BSC}_\delta)$]
\label{lem:bscUpper}
    We have that 
    
    \begin{equation*}
        \eta \ = \ \eta(\text{BSC}_\delta) \ \leq \ (1-2\delta)^2. 
    \end{equation*}
\end{lemma}

\begin{proof}
    For the BSC, we have

    \begin{equation}
        H^2(P_0,P_1) \ = \ 2(\sqrt{\delta} - \sqrt{1-\delta})^2 \ = \ 2 \ - \ 4\sqrt{\delta(1-\delta)}.
    \end{equation}

    Thus by Theorem \ref{thm:hellingerbound}, we have

    \begin{align}
    \begin{split}
        \eta \ &\leq \ 2 \ - \ 4\sqrt{\delta(1-\delta)} \ - \ \frac{(2 - 4\sqrt{\delta(1-\delta)})^2}{4} \\
        &= 1 \ - \ 4\delta (1-\delta) \\
        &= (1-2\delta)^2        
    \end{split}
    \end{align}

    as required.
\end{proof}

\begin{lemma}[Computing $\eta(\text{BSC}_\delta)$]
\label{lem:bscContraction}
    \begin{equation*}
        \eta = \eta(\text{BSC}_\delta) = (1-2\delta)^2.
    \end{equation*}
\end{lemma}

\begin{proof}
    Using Lemma \ref{lem:bscUpper}, it remains to show the lower bound. We will use result (b) in Theorem \ref{thm:sdpiMutualInformation}. Let $U\sim$ Bernoulli($\frac{1}{2}$), and $X|U\sim$ Bernoulli($\alpha$). We have

    \begin{equation}
    \label{eqn:toBeLopped}
        \eta \ \geq \ \frac{I(U;Y)}{I(U;X)} \ = \ \frac{H(U) \ - \ H(U|Y)}{H(U) \ - \ H(U|X)} \ = \ \frac{1 \ - \ H(\alpha(1-\delta)+\delta(1-\alpha))}{1 \ - \ H(\alpha)}
    \end{equation}

    by the identity for mutual information in terms of conditional entropy (\ref{eqn:condEntropyAndMutualInfo}). Observe that the derivative of $H(\alpha)$ w.r.t $\alpha$ is

    \begin{equation}
        H^\prime(\alpha) \ = \ \log \frac{\alpha}{1-\alpha}.
    \end{equation}

    Applying L'H\^{o}pital's rule twice to (\ref{eqn:toBeLopped}) yields

    \begin{equation}
        \frac{I(U;Y)}{I(U;X)} \ \to \ (1-2\delta)^2, \ \ \ \alpha \to \frac{1}{2}.
    \end{equation}

    It follows that $\eta = (1-2\delta)^2$ as required.
    
\end{proof}

\section{Evans-Schulman and the Contraction of Mutual Information in Bayesian Networks}
\label{sec:pol}

In this section we will prove the following theorem of Polyanskiy and Wu on the contraction of mutual information in Bayesian networks, from which Evans and Schulman's upper bound on $\delta^*$ in Theorem \ref{thm:es99} follows without much trouble.

\begin{definition}[Bayesian Network]
\label{def:bayesianNetwork}
    Let $G=(V,E)$ be a DAG and let $X=(X_v)$, $v\in V$ be a set of random variables indexed by $V$. We say $X$ is a \emph{Bayesian network} w.r.t. $G$ if its joint probability density function can be expressed as a product of the density functions of the individual variables conditional on their \emph{parent variables}, where pa($v$) is the set of parents of $v$, i.e. the vertices adjacent and pointing directly to $v$.

    In other words,

    \begin{equation*}
        P_X \ = \ \prod_{v\in V} P_{X_v | X_{\pa (v)}},
    \end{equation*}

    and each variable is independent of its non-descendants conditional on its parent variables, where a \emph{descendant} of $v$ is a vertex you can reach from $v$ by following directed edges.
\end{definition}

There is a particular form of Bayesian network we are interested in, as we will reduce Theorem \ref{thm:pw} to working with that form. 

\begin{figure}
    \centering
    \begin{tikzpicture}
        \node[latent] (U) {$U$};
        \node[latent, right=of U] (X_0) {$X_0$};
        \node[latent, right=of X_0] (B) {$B$};
        \node[latent, below=1.5cm of B] (A) {$A$};
        \node[latent, right=of B] (W) {$W$};
        
        \edge{B}{A};
        \edge{B}{W};
        \edge{X_0}{B};
        \edge{X_0}{A};
        \edge{U}{X_0};
        \edge{A}{W}
    \end{tikzpicture}
    \caption{Canonical Bayesian network.}
    \label{fig:canonicalBayesianNetwork}
\end{figure}
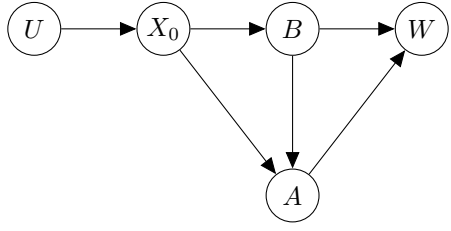

\begin{definition}[Canonical Bayesian Network]
    Suppose we have a graph $G=(V,E)$ as in Figure \ref{fig:canonicalBayesianNetwork} with the set of random variables $\{U,X_0,A,B,W\}$ satisfying

    \begin{equation*}
        P_{U,X_0,A,B, W} \ = \ P_U P_{X_0 |U} P_{B|X_0} P_{A|B,X_0} P_{W|A,B}.
    \end{equation*}
    
    We call it the \emph{canonical Bayesian network}.
\end{definition}

We have the following key result about the canonical Bayesian network.

\begin{lemma}
\label{lem:keyPolWu}
    For the canonical Bayesian network, define $\eta = \eta\left(P_{W|A,B}\right)$. We have

    \begin{equation*}
        \eta\left(P_{W,B|X_0}\right) \ \leq \ \eta \cdot \eta\left(P_{A,B|X_0}\right) \ + \ (1-\eta) \cdot \eta\left(P_{B|X_0}\right).
    \end{equation*}
\end{lemma}

\begin{proof}
    Observe that $G$ forms a Markov chain $U\to X_0 \to (A,B) \to W$, yielding the factorization of the joint distribution

    \begin{equation}
        P_{U,X_0,A,B,W} \ = \ P_U P_{X_0 |U} P_{A,B|X_0} P_{W|A,B}.
    \end{equation}

    Since $X_0$ is independent of $W$ conditional on $(A,B)$, we have that $U\to X_0 \to A \to W$ is a Markov chain conditional on $B=b$. By the SDPI Theorem \ref{thm:sdpiMutualInformation}, this Markov chain conditional on $B=b$ yields

    \begin{equation}
        I(U;W | B=b) \ \leq \ \eta I(U;A|B=b).
    \end{equation}

    Taking the expectation over $B$ and adding $I(U;B)$ to both sides yields

    \begin{equation}
        I(U;W,B) \ \leq \ \eta I(U;A,B) \ + \ (1-\eta)I(U;B)
    \end{equation}

    by (\ref{eqn:chainrule}), the chain rule for mutual information. Dividing through by $I(U;X_0)$ and taking the supremum of the LHS w.r.t. $P_{U,X_0}$ yields

    \begin{equation}
        \eta\left(P_{W,B|X_0}\right) \ \leq \ \eta \cdot \eta\left(P_{A,B|X_0}\right) \ + \ (1-\eta) \cdot \eta\left(P_{B|X_0}\right).
    \end{equation}
\end{proof}

We now introduce one model of percolation theory in a Bayesian network. It is helpful to think about $P_{X_v |X_{\pa (v)}}$ as a noisy channel.

\begin{definition}[Percolation in a Bayesian Network]
\label{def:perc}
    For DAG $G=(V,E)$, let a vertex $v\in V$ be \emph{open} with probability

    \begin{equation*}
        p(v) \ := \ \eta \left(P_{X_v |X_{\pa (v)}}\right).
    \end{equation*}

    Under this model, for two subsets $T,S \subset V$, define the \emph{percolation} probability from $T$ to $S$ by

    \begin{equation*}
        \perc (T \to S) \ := \ \Prob (\text{there exists open path from } T\to S),
    \end{equation*}

    where an \emph{open path} $T$ to $S$ is a sequence of distinct forward-facing edges sequentially joining distinct vertices $(v_1,\dots,v_k)$, where $v_1\in T$, $v_k \in S$, and $v_i$ is open for all $i=1,\dots,k$.
\end{definition}

\begin{remark}
    By thinking about $P_{X_v |X_{\pa (v)}}$ as a noisy channel, we can interpret this model as saying ``the more information we lose when passing the values of $\pa (v)$ through the channel $P_{X_v |X_{\pa (v)}}$, the less likely $v$ is to be open''.
\end{remark}

For Bayesian network $X=(X_v)$, $v\in V$ w.r.t. $G=(V,E)$, let $S\subset V$. Let $0\in V$ be a source node. We are interested in obtaining $P_{X_S |X_0}$. We can do this by a process of \emph{stitching together} the pre-defined channels $P_{X_v |X_{\pa (v)}}$, which we describe as follows.

\begin{itemize}
    \item If $v\in V$ is not a descendant of $0$, we just have that $P_{X_v |X_0} = P_{X_v}$ where $P_{X_v}$ is the marginal distribution of $X_v$.

    \item If $v\in V$ is a descendant of $0$, we can stitch the channel $P_{X_v |X_{\pa (v)}}$ onto $P_{X_{\pa (v)} |X_0}$ to find $P_{X_v | X_0}$. We can assume that $P_{X_{\pa (v)} |X_0}$ is known, as for each $u\in \pa(v)$, $P_{X_u | X_0}$ is either known through $u$ being a non-descendant, or through induction on the length of the shortest path from $0$ to $v$. All $u\in \pa (v)$ have strictly shorter paths to $0$ than $v$ does, so we can assume $P_{X_u | X_0}$. The base case for paths of length 1 is also immediate, as $0$ is just a parent.
\end{itemize}

\begin{theorem}[Contraction of Mutual Information in Bayesian Networks, Theorem 5 in \cite{PW17}]
\label{thm:pw}
    Let $G=(V,E)$ be a DAG with just one node with out-degree zero, namely $w\in V$, let and $X=(X_v)$, $v\in V$ be a Bayesian network w.r.t $G$. Let 0 be a source node in $V$. Let $S \subset V$, and define $P_{X_S |X_0}$ by stitching together channels. Then we have

    \begin{equation*}
        \eta\left(P_{X_S | X_0}\right) \ \leq \ \emph{\perc} (0\to S).
    \end{equation*}
\end{theorem}

\begin{remark}
    Suppose we have some arbitrary r.v. $U\to X_0$ prepended to $X_0$. Theorem \ref{thm:pw} provides an upper bound on the contraction coefficient that quantifies the loss of information about $U$ when observing $X_S$ compared to when observing $X_0$.
\end{remark}

\begin{proof}
    We introduce the notation $\eta(T) := \eta\left(P_{X_T | X_0}\right)$ and $\eta_v = \eta \left(P_{X_v |X_{\pa (v)}}\right)$. We will use proof by induction on $|V(G)|$, also known as the size of G. For $|V(G)|=1$, we must have either $S=\emptyset$ or $S=\{X_0\}$. In the former case, both sides of the statement we wish to prove are trivially 0, and in the latter case, both sides are trivially 1.

    Suppose the statement is true for all graphs smaller than $G$. If $w \notin S$, then we can exclude it from $G$ and we are done by the induction hypothesis. Otherwise, define $S_A = \pa (w) \setminus S$ and $S_B = S \setminus w$. Correspondingly define random variables $A=X_{S_A}$, $B=X_{S_B}$, $W = X_w$.

    We have that $0\notin A$ w.l.o.g., as if $0\in A$, we can replace the vertex with a fake vertex $0^\prime$ and $X_{0^\prime}=X_0$, and then move 0 out of $A$.
    
    Prepending arbitrary $U\to X_0$ the graph, we have that the joint distribution of $(U,X_0,A,B, W)$ can be factorised as

    \begin{equation}
        P_{U,X_0,A,B, W} \ = \ P_U P_{X_0 |U} P_{B|X_0} P_{A|B,X_0} P_{W|A,B}
    \end{equation}

    where $P_{B|X_0}$ is found by stitching, and $P_{A|B,X_0}$ is also computed by stitching once $P_{B|X_0}$ is found. We can easily find $P_{W|A,B}$ since $\pa(w) \subset S_A \cup S_B$. Finding $P_U$, $P_{X_0|U}$ is immediate from how $U$ and $X_0$ are defined.
    
    This means that random variables $(U,X_0,A,B, W)$ form the canonical Bayesian network, and we can apply Lemma \ref{lem:keyPolWu} to immediately yield (note our new notation)

    \begin{equation}
        \eta (S) \ \leq \ \eta_w \eta \left(S_A \cup S_B\right) \ + \ (1-\eta_w) \eta \left(S_B\right).
    \end{equation}

    Since $S_A$ and $S_B$ live entirely on $G\setminus \{w\}$, we have by induction that

    \begin{equation}
        \eta (S) \ \leq \ \eta_w \perc\left(0\to S_A \cup S_B\right) \ + \ (1-\eta_w) \perc\left(0\to S_B\right).
    \end{equation} 

    By Definition \ref{def:perc}, we have that $p(w) = \eta_w$. We thus have

    \begin{align}
    \begin{split}
        \perc(0\to S) \ &= \ \perc\left(0 \to S_B \cup \{w\}\right) \\
        &= \ \perc\left(0 \to S_B\right) \ + \ \perc\left(0 \to \{w\}\right) \\
        & \ \ \ \ - \ \Prob \left(\exists \text{ open path } 0\to S_B \text{ and } \exists \text{ open path } 0\to \{w\}\right) \\
        &= \ \perc\left(0 \to S_B\right) \ + \ p(w)\perc\left(0 \to S_A \cup S_B\right) \ - \ p(w)\perc\left(0 \to S_B\right) \\
        &= \ \eta_w \perc\left(0\to S_A \cup S_B\right) \ + \ (1-\eta_w) \perc\left(0\to S_B\right) \\
        &\geq \ \eta(S)
    \end{split}
    \end{align}

    as required.
\end{proof}

We will now head towards proving Evans-Schulman's 1999 upper bound on the threshold noise for reliable computation by fan-in $k$ noisy circuits (Theorem \ref{thm:es99}). The key observation is that the DAG underlying a Boolean circuit also underlies a corresponding Bayesian network. We also need the following definition and theorem:

\begin{definition}
\label{def:karyTree}
    A \emph{$k$-ary tree} is a directed graph with a root vertex $u$ such that for any other vertex $v$ there is precisely one path from $u$ to $v$ and each vertex has no more than $k$ children.
\end{definition}

\begin{theorem}[Contraction of Mutual Information Between an Input and the Output, Theorem 33.2 in \cite{PWTB}.]
\label{thm:contractionOfMutualInformationWithDepth}
    For an $n$-input $\delta$-noisy Boolean circuit with fan-in $k$, let the inputs be binary random variables $X_1, \dots, X_n$, and the output be binary random variable $Y$. Let $d_i$ be the length of the shortest path between $X_i$ and $Y$ for $i=1,\dots,n$.

    Provided $k(1-2\delta)^2 \leq 1$, we have

    \begin{equation*}
        I(X_i;Y) \ \leq \ \left(k(1-2\delta)^2\right)^{d_i}.
    \end{equation*}
\end{theorem}

\begin{figure}
    \centering
    \begin{minipage}[t]{0.45\textwidth}
        \centering
        \begin{tikzpicture}[scale=1.2]
            \node[draw] (1) at (0,4) {$X_1$};
            \node[draw] (2) at (1,4) {$X_2$};
            \node[draw] (3) at (2,4) {$X_3$};
            \node[draw] (4) at (3,4) {$X_4$};
            \node[draw] (5) at (4,4) {$X_5$};
            \node[draw] (a) at (1,3) {$G_1$};
            \node[draw] (b) at (2.5,3) {$G_2$};
            \node[draw] (c) at (1.75,2) {$G_3$};
            \node[draw] (d) at (3.5,2) {$G_4$};
            \node[draw] (e) at (2.625,1) {$G_5$};
            \node[] (f) at (2.625,0) {$Y$};
            \draw[->] (1) -- (a);
            \draw[->] (2) -- (a);
            \draw[->] (3) -- (a);
            \draw[->] (3) -- (b);
            \draw[->] (4) -- (b);
            \draw[->] (a) -- (c);
            \draw[->] (3) -- (c);
            \draw[->] (b) -- (c);
            \draw[->] (b) -- (d);
            \draw[->] (5) -- (d);
            \draw[->] (c) -- (e);
            \draw[->] (d) -- (e);
            \draw[->] (e) -- (f);
        \end{tikzpicture}
    \end{minipage}\hfill
    \begin{minipage}[t]{0.45\textwidth}
        \centering
        \begin{tikzpicture}[scale=0.8]
            \node[draw, circle] (1) at (0,6) {};
            \node[draw, circle] (2) at (1,6) {};
            \node[fill=red, draw, circle] (3) at (2,6) {};
            \node[fill=green, draw, circle] (4) at (4,6) {};
            \node[draw, circle] (5) at (5,6) {};
            \node[fill=blue, draw, circle] (6) at (6,6) {};
            \node[draw, circle] (7) at (7,6) {};
            \node[draw, circle] (8) at (1.5,4) {};
            \node[fill, draw, circle] (9) at (3,4) {};
            \node[draw, circle] (10) at (4.5,4) {};
            \node[draw, circle] (11) at (6.5,4) {};
            \node[draw, circle] (12) at (8.5,4) {};
            \node[draw, circle] (13) at (3,2) {};
            \node[draw, circle] (14) at (7.5,2) {};
            \node[draw, circle] (15) at (5.25,0) {Y};
            \draw[->] (15) -- (13);
            \draw[->] (15) -- (14);
            \draw[->] (13) -- (8);
            \draw[->] (13) -- (9);
            \draw[->] (13) -- (10);
            \draw[->] (14) -- (11);
            \draw[->] (14) -- (12);
            \draw[->] (8) -- (1);
            \draw[->] (8) -- (2);
            \draw[->] (8) -- (3);
            \draw[->] (10) -- (4);
            \draw[->] (10) -- (5);
            \draw[->] (11) -- (6);
            \draw[->] (11) -- (7);
        \end{tikzpicture}
    \end{minipage}
    \caption{The four paths from $X_3\to Y$ in the circuit are encoded as coloured vertices in the $k$-ary tree. Writing the paths as the sequence of distinct vertices traversed, red is $(X_3,G_1,G_3,G_5)$, green is $(X_3,G_2,G_3,G_5)$, blue is $(X_3,G_2,G_4,G_5)$, black is $(X_3,G_3,G_5)$.}
    \label{fig:karytreesandcircuits}
\end{figure}
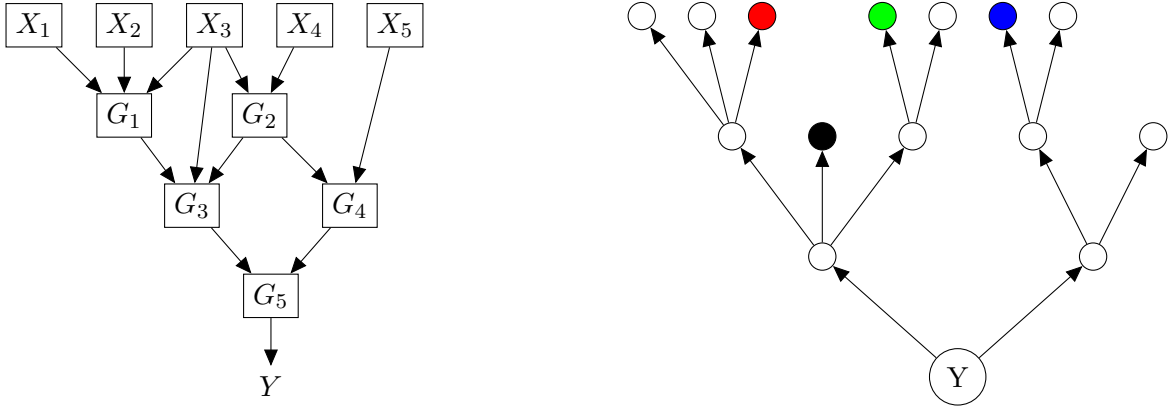

\begin{proof}
    First observe that a $\delta$-noisy Boolean circuit with underlying DAG $G=(V,E)$ has a corresponding Bayesian network $(X_v)$. For this Bayesian network, we have that for all $v\in V$ with in-degree greater than zero,

    \begin{equation}
        P_{X_v | X_{\pa (v)}} \ = \ \text{BSC}_\delta
    \end{equation}

    by the definition \ref{def:noisyCircuits} of a noisy circuit.

    Thus by Lemma \ref{lem:bscContraction}, we have that for all $v\in V$ with in-degree greater than zero,

    \begin{equation}
        p(v) \ = \ \eta_v \ = \ (1-2\delta)^2.
    \end{equation}

    Letting $U=X_i$ and applying Theorem \ref{thm:sdpiMutualInformation} (the SDPI for mutual information) and (\ref{eqn:entropyAndMutualInfo}), we have

    \begin{equation}
        I(X_i;Y) \ \leq \ \eta\left( P_{Y|X_i} \right) H(X_i) \ \leq \ \eta\left( P_{Y|X_i} \right).
    \end{equation}

    By Theorem \ref{thm:pw} we have the upper bound 
    
    \begin{equation}
        \eta\left( P_{Y|X_i} \right) \ \leq \ \perc \left(X_i \to Y\right).
    \end{equation}

    We can upper bound $\perc \left(X_i \to Y\right)$ in the following way. Let $\pi$ be a path from $X_i$ to $Y$, and let $\pi$ have length $l(\pi)$. The probability this path is open is equal to $(1-2\delta)^{2l(\pi)}$. Thus we can sum over all paths and get

    \begin{equation}
        \perc \left(X_i \to Y\right) \ \leq \ \sum_{\pi : X_i \to Y} (1-2\delta)^{2l(\pi)}.
    \end{equation}

    We wish to show that $\sum_{\pi : X_i \to Y} (1-2\delta)^{2l(\pi)} \leq \left(k(1-2\delta)^2\right)^{d_i}$ when $k(1-2\delta)^2 \leq 1$. We can represent paths $X_i\to Y$ as vertices of a $k$-ary tree with root $Y$ by working backwards. Starting from $Y$, encode the path to $X_i$ one edge at a time. Represent the different inputs the path could take at each gate as moving to a different child in the $k$-ary tree, and do this for each edge in the path until we reach $X_i$. The vertex reached in the $k$-ary tree encodes the path. See Figure \ref{fig:karytreesandcircuits} for an example.

    In the $k$-ary tree, let the vertices corresponding to paths from $X_i\to Y$ in the circuit be the set $T$. Observe that (a) the minimum depth (distance from the root) of any $t\in T$ is $d_i$ by definition, and also observe that (b) no vertex in $T$ can be the descendant of another vertex in $T$. 
    
    Suppose $T$ were a set of $k^{d_i}$ vertices at depth $d_i$. Then

    \begin{equation}
        \sum_{t\in T} (1-2\delta)^{2\cdot\text{depth}(t)} \ = \ \left(k(1-2\delta)^2\right)^{d_i}.
    \end{equation}

    Provided that $k(1-2\delta)^2 \leq 1$, there is no way to change $T$ that increases the value of $\sum_{t\in T} (1-2\delta)^{2\cdot\text{depth}(t)}$. The maximum number of vertices with depth $d_i$ in a $k$-ary tree is $k^{d_i}$, so we can only add vertices with depth greater than $d_i$ due to (a). However, we must remove a vertex with depth $d_i$ to add at most $k$ vertices with depth $d_i +1$ due to (b).

    Due to the condition $k(1-2\delta)^2 \leq 1$, this change always decreases $\sum_{t\in T} (1-2\delta)^{2\cdot\text{depth}(t)}$. It is easy to see that replacing a vertex in $T$ with any choice of descendants also decreases $\sum_{t\in T} (1-2\delta)^{2\cdot\text{depth}(t)}$. The extra vertices do not compensate for the larger powers on  $(1-2\delta)^2$. 

    Thus provided $k(1-2\delta)^2 \leq 1$, we indeed have

    \begin{equation}
        \sum_{\pi : X_i \to Y} (1-2\delta)^{2l(\pi)} \ \leq \ \left(k(1-2\delta)^2\right)^{d_i}.
    \end{equation}

    Together, our upper bounds imply

    \begin{equation}
    \label{eqn:basicallyEvansSchulman}
        I(X_i;Y) \ \leq \ \left(k(1-2\delta)^2\right)^{d_i}.
    \end{equation}
    
    when $k(1-2\delta)^2 \leq 1$, as desired.
\end{proof}

We are now in a position to prove Evans-Schulman's 1999 result.

\begin{proof}[Proof of Theorem \ref{thm:es99}]

    Consider the Boolean function
    
    \newcommand{\XOR}{\text{XOR}}
    \begin{equation}
        \XOR\left(X_1,\dots,X_n\right) \ = \ \begin{cases}
		1, & \text{if \# $X_i=1$ is odd}\\
            0, & \text{otherwise.}
	\end{cases}
    \end{equation}

    This function depends on each one of its inputs, so if it were to be reliably computed by a noisy Boolean circuit $C$ with $Y=C(X_1,\dots,X_n)$, we certainly require

    \begin{equation}
        I(X_i;Y) \ > \ 0
    \end{equation}

    for all $i=1,\dots,n$. Observe that there must be at least one input $X_i$ with

    \begin{equation}
    \label{eqn:dilowerBound}
        d_i \ \geq \ \frac{\log n}{\log k},
    \end{equation}

    as if not, let $d$ be the depth of the circuit. We must have $d < \frac{\log n}{\log k}$, implying $k^d < n$, which is a contradiction as a fan-in $k$ circuit with depth $d$ (and a single output) has at most $k^d$ inputs.

    Thus taking $n\to \infty$ takes $d_i\to\infty$ for some input $X_i$. But then by Theorem \ref{thm:contractionOfMutualInformationWithDepth}, when $k(1-2\delta)^2 < 1$, we have for some $i\in\{1,\dots,n\}$ that

    \begin{equation}
        I(X_i;Y) \ \to \ 0, \ \ d_i \ \to \ 0.
    \end{equation}

    Thus we cannot have reliable computation for all Boolean functions with fan-in $k$ $\delta$-noisy circuits (as XOR is a counterexample) unless

    \begin{equation}
        k(1-2\delta)^2 \ \geq \ 1.
    \end{equation}

    Rearranging, we find we require

    \begin{equation}
        \delta \ \leq \ \delta^*_{\text{ES}} \ = \ \frac{1}{2} \ - \ \frac{1}{2\sqrt{k}} 
    \end{equation}

    is a necessary condition for reliable computation. Thus

    \begin{equation}
        \delta^* \ \leq \ \delta^*_{\text{ES}}
    \end{equation}

    as required.
\end{proof}

\section{Conclusion}

In this essay we have introduced the field of reliable computation and provided a proof that all Boolean functions can be reliably computed by a noisy circuit consisting of 3-input minority gates. 

We have also used information theoretic techniques such as the Strong Data-Processing Inequality to derive the best known upper bound $\delta^* \leq \delta^*_{\text{ES}}$ on the threshold noise for which reliable computation is possible by a fan-in $k$ circuit.

Recall that the threshold noise for reliable computation by formulas of fan-in $k$, $\delta^*_f (k)$, is known for all odd $k$, and is a lower bound for $\delta^* (k)$. We discussed that $\delta^*_f (k) \leq \delta^*(k) \leq \delta^*_{\text{ES}}(k)$ is an asymptotically tight bound for large $k$.

Finding $\delta^* (k)$ for any value of $k$ remains an open question. Another open question is finding $\delta^*_f (k)$ for even $k>2$. In fact, even finding $\delta^*_{\text{3MAJ}}$ is open, despite reliable computation with 3MAJ gates being first proven possible in 1952.

It is reasonable to ask why we care. While information theory is an abstraction, the key quantities such as entropy and mutual information can be understood intuitively. von Neumann did not use any information theoretic techniques in his proof of reliable computation via 3MAJ gates. Yet he nonetheless felt it appropriate to include a brief introduction to information theory in the series of lectures where he first introduced the topic of reliable computation \cite{V52}.

von Neumann's original source of interest was understanding how the brain works. He felt that this area of mathematical study would be intrinsically linked to information theory, and he was correct. Perhaps he was also correct about an implication of improving our understanding of reliable computation being further understanding of our own brains. This could in turn have positive consequences such as improving teaching and healthcare, or perhaps even more murky consequences, such as simulating a human conscience.

\end{document}